\documentclass[apjl]{emulateapj}
\usepackage{apjfonts}
\submitted{}

\def\hi{\ifmmode {\rm H}\,{\sc i}~ \else H\,{\sc i}~\fi}
\def\kms{\rm\,km\,s^{-1}}

\def\chandra {{\it Chandra}}
\def\xmm {{\it XMM}}
\def\xmmnewton {{\it XMM--Newton}}
\def\fuse {{\it FUSE}}
\def\cv {\ion{C}{5}}
\def\cvi {\ion{C}{6}}

\def\neix {\ion{Ne}{9}}
\def\nex {\ion{Ne}{10}}

\def\ovi {\ion{O}{6}}
\def\ovii {\ion{O}{7}}
\def\oviii {\ion{O}{8}}

\def\kms    {~km~s$^{-1}$}

\slugcomment{Submitted to ApJ Letters}

\shorttitle{X--ray absorption in AGN host galaxies}
\shortauthors{Williams et al.}

\begin{document}

\title{On the nature of the $z=0$ X-ray absorbers: II. The contrast between 
local and AGN host galaxy absorption }

\author{Rik J. Williams\altaffilmark{1}, 
        Smita Mathur\altaffilmark{2},
	Fabrizio Nicastro\altaffilmark{3,4}}
\altaffiltext{1}{Leiden Observatory, Leiden University, PO Box 9513,
                 2300 RA Leiden, The Netherlands}
\altaffiltext{2}{Department of Astronomy, The Ohio State University,
                 Columbus, OH, 43210, USA}
\altaffiltext{3}{Osservatorio Astronomico di Roma, Istituto Nazionale di 
                 AstroFisica, Italy}
\altaffiltext{4}{Harvard--Smithsonian Center for Astrophysics, Cambridge,
                 MA, 01238, USA}
\email{williams@strw.leidenuniv.nl}

\begin{abstract}
We search for highly--ionized gas near three AGN host galaxies
using the \chandra\ low--energy transmission grating spectrograph.
Strong absorption lines from such gas are seen at $z=0$, most likely
from one or more of the following components: (1) a Galactic corona,
(2) the Local Group medium, and (3) an extended warm--hot intergalactic
medium (WHIM) filament passing through our local overdensity.  
Since AGNs reside within host galaxies that
are also expected to sit within cosmically overdense regions, similar
absorption resulting from these three
components should appear at the AGN redshifts as well.
However, no such absorption is seen.  The lack of strong absorption
lines is likely a result of the gas in these host galaxies and surrounding
galaxy clusters being much hotter, and hence more highly ionized, than 
the gas in the Local Group$+$Galaxy system.  We conclude that
WHIM filaments produce no measurable absorption lines at the AGN 
redshifts, and therefore contribute at most a small fraction of the 
observed $z=0$ warm--hot gas.
\end{abstract}


\section{Introduction}
The study of local warm--hot gas has been revolutionized by the
advent of sensitive space--based observatories such as \chandra,
\xmmnewton, and the {\it Far Ultraviolet Spectroscopic Observer} (\fuse).
By taking spectra of bright background quasars, absorption lines from
highly--ionized heavy elements at $z\sim 0$ can be seen in nearly all
directions in both the X--ray
\citep{nicastro02,williams05,williams06b,williams07} and
far--UV \citep{wakker03}.  These lines are thought
to trace a variety of mass components; while strong, broad \ovi\ absorption
is associated with the Galactic thick disk \citep{savage03},
high--velocity \ovi\ clouds more likely reside in the extended Galactic
halo, with a subpopulation of these showing evidence of extragalactic
(Local Group) kinematics \citep{sembachetal03,nicastro03}.
More enigmatic still are the X--ray absorbers, which exhibit kinematics and
column densities inconsistent with both the low-- and high--velocity
\ovi\ absorbers \citep[e.g.,][]{williams05}.  Whether these absorbers
arise primarily from a Galactic or an extragalactic medium is still a
matter of debate.

In the ``extragalactic'' picture, the local X--ray absorption arises
primarily from the intragroup medium filling the Local Group (analogous
to the extended X--ray emitting gas seen in massive galaxy clusters), or
from the larger--scale IGM filament(s) connecting the overdensities traced by the Local
Group and nearby structures such as the Virgo cluster 
\citep{kravtsov02}.  Such structure, containing most of the baryonic
mass at low redshifts, is now a generic prediction of
cosmological simulations and has come to be known as the warm--hot
intergalactic medium \citep[WHIM;][]{cen99,dave01}.  On the other hand, much
of the local X--ray absorption may also be within a few tens of kpc, resulting 
from a hot Galactic corona or fountain driven by supernova winds
\citep{sembach03,collins05,wang05}.

In a companion paper to this one \citep{mathur07}, we propose that
the X--ray absorption lines seen at $z=0$ are likely to contain contributions
from both Galactic and intra--Local Group media.  Additional contributions
could come from the local large--scale intergalactic
filament, and perhaps a supernova remnant lying near the occasional line of 
sight \citep[e.g., 3C~273;][]{savage93}.  All of these components would
apparently be blended into a single line within 
\chandra's $\sim 700$\kms\ resolution.  While it appears that
poor groups of galaxies similar to our own can produce the observed 
\ovii\ column densities \citep{mathur07}, the relative
contributions from Galactic and extragalactic phenomena in our own Local Group 
are as yet undetermined.

The intervening WHIM lines found by \citet{nicastro05} are quite weak
relative to the local absorption; indeed, intervening absorption systems
of comparable column density have yet to be discovered.
Thus, the strength of the $z=0$ absorption is doubtless related to our
peering out from a ``special'' vantage point centered on a dense region ---
in the disk
of a spiral galaxy within a poor galaxy group.  However, inherent to
every quasar spectrum is one other such special location: the AGN host
galaxy.  As AGNs typically occur at the centers of massive galaxies,
some within groups and clusters, large column densities of warm--hot gas 
should also be present at the quasar redshifts as at
$z=0$.  These components could include intra-- and intergalactic gas 
surrounding the host galaxy,
plus gas directly related to the accretion process (such as a 
warm absorber or outflow).  In other words, if the Galaxy is a typical
(non--AGN) system, absorption column densities
near AGN host galaxies should be {\it at least} as strong as those
observed locally, and significant discrepancies (vis--\`a--vis differences
between the host galaxies' and the Milky Way's systems) can provide clues
to the nature of the local absorption.
In this {\it Letter} we search for absorption lines at the redshifts
of three AGNs observed with the \chandra\ grating spectrograph, and contrast
these with the absorption lines seen at $z=0$ in the same spectra.  

\begin{deluxetable*}{lccccccc}
\tabletypesize{\footnotesize}
\tablecolumns{8}
\tablewidth{450pt}
\tablecaption{Absorption lines at $z=0$ and $z_{\rm host}$ \label{tab_lines}}
\tablehead{
\colhead{Object} &
\colhead{$z_{\rm abs}$}   &
\colhead{$W_\lambda$\tablenotemark{a}} &
\colhead{$W_\lambda$\tablenotemark{a}} &
\colhead{$W_\lambda$\tablenotemark{a}} &
\colhead{$W_\lambda$\tablenotemark{a}} &
\colhead{$W_\lambda$\tablenotemark{a}} &
\colhead{$W_\lambda$\tablenotemark{a}} 
\\
\colhead{} &
\colhead{} &
\colhead{\nex} &
\colhead{\neix} &
\colhead{\oviii} &
\colhead{\ovii} &
\colhead{\cvi} &
\colhead{\cv}
\\
\colhead{} &
\colhead{} &
\colhead{12.13\,\AA} &
\colhead{13.45\,\AA} &
\colhead{18.97\,\AA} &
\colhead{21.60\,\AA} &
\colhead{33.74\,\AA} &
\colhead{40.27\,\AA} 
}

\startdata
Local \\
\hline
Mrk 421       &0 &\nodata &$1.6\pm 0.5$ &$1.7\pm 0.6$ &$11.4\pm 0.8$ &$7.4\pm 1.5$ &$15.7\pm 3.3$\\
PKS 2155--304 &0 &\nodata &$3.8\pm 1.2$ &$6.2\pm 1.4$ &$12.2^{+3.7}_{-2.8}$ &$5.5\pm 2.6$ &$9.4^{+7.1}_{-5.2}$\\
3C 273        &0 &\nodata &$6.9\pm 2.0$ &$7.4\pm 2.9$ &$26.5\pm 3.1$ &$19.5\pm 8.2$ &$50\pm 20$\\
Mrk 279\tablenotemark{b} &0 &\nodata &\nodata &\nodata &$26.6\pm 6.2$ &\nodata &\nodata\\
\hline\hline
Host galaxies \\
\hline
Mrk 421       &0.030 &$<1.4$ &$<3.3$ &$<1.8$  &$<3.8$  &$<5.4$  &$<12.9$\\
PKS 2155--304 &0.116 &$<5.4$ &$<5.8$ &$<7.3$  &$<6.8$  &$<12.6$ &$<8.6$ \\
3C 273        &0.158 &$<7.2$ &$<4.1$ &$<11.7$ &$<14.4$ &$<47.9$ &$<24.9$\\
\enddata
\tablenotetext{a}{All equivalent widths are quoted in m\AA\ with $1\sigma$
errors; upper limits are $2\sigma$ confidence.}
\tablenotetext{b}{Redshifted absorption lines are not measured for this
object, due to absorption from a strong AGN outflow.}
\end{deluxetable*}

\section{Data Reduction and Measurements}
In this analysis we focus on data from the \chandra\ Low--Energy Transmission
Grating (LETG) coupled with its two standard readout detectors, the
High Resolution Camera and AXAF CCD Imaging Spectrometer spectroscopic
arrays (HRC--S and ACIS--S respectively).  While the {\it XMM--Newton} 
Reflection Grating Spectrometers (RGS1/RGS2) together have comparable 
sensitivity
to the \chandra\ LETG, the RGS contains many narrow detector features and
a more restricted wavelength range \citep{williams06a}. Furthermore,
there is evidence for significant cross--calibration uncertainties between 
\xmm/RGS and \chandra/LETG \citep{rasmussen07}; thus, for the sake of 
consistency we will focus solely on the \chandra\ data for this analysis.

Local ($z\sim 0$) absorption has been confidently detected by \chandra\ in 
four local AGN sight lines: Mrk 421 \citep{williams05}, Mrk 279 
\citep{williams06b}, PKS 2155--304 \citep{nicastro02,williams07},
and 3C 273 \citep{fang03}.  The far--UV and X--ray spectra of Mrk 279 indicate
the presence of strong intrinsic absorption from an AGN outflow 
\citep{fields07}; thus, only
the $z=0$ absorption seen in this spectrum will be considered in this analysis
All \chandra/LETG observations of these four sources employing the ACIS--S
and HRC--S readout detectors were downloaded from the \chandra\ archive
and fully re--reduced with CIAO 
4.0b1\footnote{\url{http://cxc.harvard.edu/ciao/}}.  Zeroth--order source
positions were checked by eye and corrected if necessary, and grating
spectra and response files for all observations generated.  The spectra
from positive and negative grating orders were added for each observation,
and all observations for the same object subsequently added.  Response
files for the positive and negative ACIS--S orders, and from orders
$-6$ through $+6$ for HRC--S\footnote{This is necessary to properly account 
for contamination from unresolved higher spectral orders; see
\url{http://cxc.harvard.edu/ciao/threads/hrcsletg\_orders/}}, were similarly
combined to produce average ACIS and HRC instrumental response functions
for each object.

Spectral continua and absorption lines were modeled with the 
fitting program 
{\it Sherpa}\footnote{\url{http://cxc.harvard.edu/sherpa/}}.
A single powerlaw plus Galactic absorption model was initially fit to each 
spectrum over the 10--50\,\AA\ range to properly determine
the level of higher--order contamination in HRC--S.  Individual 
lines were fitted with negative Gaussians in $\sim 2$\,\AA\ windows around each
wavelength of interest, retaining the continuum shape found with the 
broadband fit but allowing the amplitude to vary.  Any residual broad
features in the spectra (much larger than the LETG line--spread function
width, $\sim 50$\,m\AA) were removed by including broad Gaussian components.
The continua were
allowed to vary independently for HRC--S and ACIS--S to allow for 
flux variations between the different observation epochs, while absorption line
amplitudes were determined using a joint fit to both instrumental
spectra.  Upper limits (at the
$2\sigma$/95\% confidence level) were determined by allowing the 
line wavelength to vary within $\pm 0.02$\AA\ of the expected central
wavelength to take into account the systematic uncertainty in the 
LETG dispersion relation.\footnote{See \url{http://cxc.harvard.edu/cal/}}

At the AGN redshifts, we measured six absorption lines that were
detected in the Mrk 421 spectrum at $z=0$, and are expected
to peak in abundance over a range of temperatures spanning roughly a factor
of ten under collisional
ionization: $T\sim 10^{6.5}$\,K
(\oviii, \nex), $10^{6}$\,K (\ovii, \cvi, \neix) and $10^{5.5}$\,K
(\cv).  Although this latter temperature is expected to occur in galactic
interstellar media \citep[e.g., as observed in the thick disk of the Galaxy
through strong \ovi\ absorbers with FUSE;][]{savage03}, it may provide
a useful indicator of whether the AGN host galaxies contain similar
warm--hot gas column densities as the Galaxy but at lower temperatures.
To ensure that consistent reduction methods and calibrations are
used throughout, we re--measured the strengths of the same absorption
lines at $z=0$ using the same broadband continuum fits. 
Table~\ref{tab_lines} shows the strengths (or upper limits)
for all measured local and redshifted absorption lines.

\begin{figure*} 
\begin{center}
\includegraphics[scale=0.5]{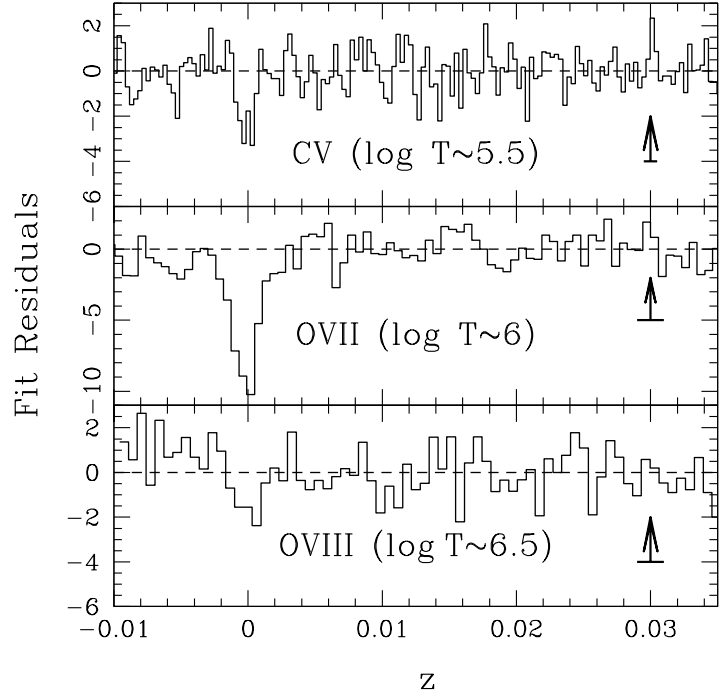}
\end{center}
\caption{\chandra--LETG spectrum of Mrk 421 at the wavelengths of three
strong absorption lines expected at various gas temperatures.  In all
cases a strong line appears at $z=0$ but nothing is seen at the host
galaxy redshift (marked with the arrows; the horizontal arrow base
shows the systematic LETG wavelength uncertainty).
\label{fig_m421res}}
\end{figure*}

As the table shows, at the AGN redshifts none of these absorption lines are 
formally detected at a significance larger than $2\sigma$; indeed, 
visual inspection of the spectra confirms the absence of any apparent
absorption at or near the AGN.  Note that \citet{rasmussen07} find
a weak ($2.6$\,m\AA), $z=0.032$ \ovii\ line at $3.7\sigma$ significance 
with \xmm; 
however, this is not seen in the \chandra\ spectrum and may be a $z=0$
\ion{O}{5} line near the same wavelength.  Figure~\ref{fig_m421res} shows the
continuum fit residuals for \oviii, \ovii, and \cv\ in the highest--quality
spectrum (Mrk 421).  Though absorption in all three atomic species is
detected at $z=0$, none is seen at $z=0.03$.  Furthermore, these three
lines can be present at temperatures ranging from $10^5-10^7$\,K; if the Mrk 421
host galaxy contained gas at a column density comparable to that observed at 
$z=0$ but with a somewhat higher or lower temperature, the strength of the
\oviii\ or \cv\ line should be boosted respectively.  However, this is 
clearly not the case.  

\begin{figure*}
\begin{center}
\includegraphics[scale=0.5]{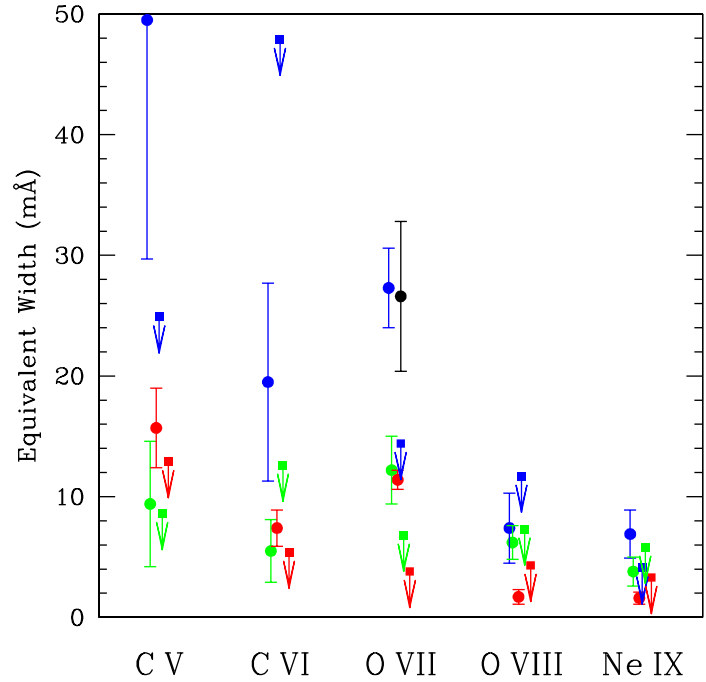}
\end{center}
\caption{Comparison of the measured equivalent widths of five $z=0$ absorption
lines seen toward the four AGNs ({\it circles with error bars}) along with
the $2\sigma$ upper limits measured at three AGN host galaxy redshifts 
({\it squares with arrows}).  Measurements and limits for 
Mrk 421 are denoted by red points, PKS2155--304 by green, 3C 273 by blue,
and Mrk 279 by black.  Only the $z=0$ \ovii\ point is plotted for Mrk 279.
\label{fig_limits}}
\end{figure*}

A visual representation of the data in Table~\ref{tab_lines} is shown in 
Figure~\ref{fig_limits}.  The discrepancy is most evident for
\ovii: most of the $2\sigma$ upper limits at the AGN redshifts are
inconsistent with the typical local absorption strength.  Even
for the other four lines shown in this figure, the upper limits
are nonetheless below or comparable to the $z=0$ measurements.  It is
thus apparent that the gaseous environments of the AGN host galaxies
are in fact dramatically different from that around the Galaxy.  

\section{AGN host galaxy properties}
Since strong $z=0$ \ovii\ absorption ($>10$\,m\AA\ in all cases) is seen toward
AGNs anytime sufficient S/N is acquired, it is unlikely that the lack of
observed host galaxy absorption is a result of a low \ovii\ covering
fraction.  Instead, this discrepancy must be due to fundamental differences
between the Milky Way and host galaxy systems themselves. A summary of 
the three host galaxy systems follows.

\subsection{Mrk 421 and PKS 2155--304}
As the two highest signal--to--noise spectra, the absence of absorption at
the redshifts of Mrk 421 and PKS 2155--304 is particularly striking.  Not
only do both of these spectra exhibit strong similarities in their $z=0$
absorption properties (consistent column densities, Doppler parameters,
and temperature constraints), but the AGN environments are also remarkably
similar.  Both are BL Lac objects contained within giant elliptical host 
galaxies \citep{ulrich75,kikuchi85,falomo91}.  Such galaxies
are known to contain large quantities of hot gas from X--ray emission
studies \citep[e.g.,][]{brighenti97}, with the most massive ellipticals
exhibiting gas temperatures of $T\sim 10^7$\,K.  At this temperature
oxygen is expected to be fully ionized.  The nuclei of both
these galaxies are powerful blazars; thus, intense X--ray flux from the
central source may further contribute to the ionization of gas in the central
regions of the galaxy, and jets or outflows from the AGN might evacuate
galactic gas from the line of sight.  Furthermore, these host galaxies
reside at the centers of clusters \citep{ulrich78,falomo93}.  Since the 
intracluster media within massive galaxy clusters 
typically exhibit similarly high temperatures ($T\ga 10^7$\,K), this cluster
environment is unlikely to produce strong absorption lines.

\subsection{3C 273}
The host galaxy of 3C 273 was studied extensively with the HST Advanced
Camera for Surveys (ACS) coronagraph by \citet{martel03}.  They find
a morphology similar to an early--type galaxy (dominated by an extended
stellar halo at large radii), but also evidence for spiral structure 
and a possible recent major merger in the central region.  Although the 
existence of a spiral and/or
merger morphology might increase the likelihood of seeing strong absorption
(due to higher star formation rates than in giant ellipticals, and hence
more supernova remnants and ``galactic fountain'' activity), it may be
that the fully ionized hot gas in the outer regions dominates the 
total column density.  Furthermore, 3C 273 is not known to be a member 
of a group or cluster, in which case we don't expect to see a contribution
from the surrounding intragroup medium like at $z=0$.

In any case, the comparatively low sensitivity of this spectrum ($2\sigma$ 
upper limit of $<14.4$\,m\AA\ on redshifted \ovii) is the primary limitation
to detecting absorption near the host galaxy.  Only two of the four $z=0$
systems show \ovii\ absorption stronger than the upper limit found for
the host galaxy, so it is perhaps not surprising that no redshifted
absorption is seen here.

\section{Discussion and Conclusions}
We searched for absorption near AGN host galaxies in the three 
highest--quality \chandra/LETG AGN spectra, but none could be seen within
the sensitivity of our observations.  This is in stark contrast to $z=0$,
where absorption of $W_\lambda$(\ovii)$\ga 10$\,m\AA\ is seen in all
directions whenever enough signal is accumulated.  Given this discrepancy,
and by comparing the characteristics of the AGN host galaxy systems
with the $z=0$ Galaxy$+$Local Group system, we propose three likely
contributing factors:
\begin{enumerate}
\item{{\it The host galaxies reside in high--mass, and thus very high--temperature systems.}  
Mrk 421, PKS 2155--304, and possibly 3C 273 are hosted
by massive ellipticals; the former two are also at the centers of 
rich galaxy groups or clusters.  Although both massive ellipticals and
clusters contain large amounts of hot gas, at $\ga 10^7$\,K even the heavier
elements analyzed herein are fully ionized.  On the other hand, the
Local Group is a relatively low--mass system and the measured gas
temperatures at $z=0$ ($\sim 10^6$\,K) are in a range that maximizes
the abundances of ions easily measured by \chandra.  
}
\item{{\it The IGM filaments surrounding the Local Group and host galaxies
contribute negligibly to the total absorption.}  The extended intergalactic
medium filaments surrounding the AGN host galaxies should in principle be 
similar
to that around our own Local Group, and at sufficient distance to be essentially
unaffected by the galaxy and/or AGN.  Thus, given the ubiquity of absorption
lines at $z=0$, any ``local filament'' contribution to these lines should 
also be visible around most AGN as well.  Since it is in fact not visible,
we conclude that large--scale filamentary IGM structures are not an 
important contributor to the $z=0$ X--ray absorption.  This is consistent
with the weakness of the absorption seen in the intervening WHIM filaments
reported by \citet{nicastro05}.  However, it stands in conflict with the
marginal detection of X-ray absorption (also presumably from a
surrounding WHIM filament) toward the Coma cluster in a shallower
\xmm\ spectrum by \citet{takei07}.
}
\item{{\it Lack of star--formation activity in the host galaxies.}  
As supernovae are known to produce large ``bubbles'' of hot gas, it stands to
reason that an occasional AGN sight line will pass near enough to one of these
supernova remnants to exhibit strong X--ray absorption.  This is likely
the cause of the strong absorption seen locally toward 3C 273 
\citep{savage93}.  Not only is the sample size
here relatively small, thereby lowering the chance probability of
a sight line intercepting a SNR in one of these host galaxies (unless the 
average SNR covering fraction is very large), but the host galaxies of these
AGN are early--type galaxies and hence contain very little star formation,
further decreasing the likelihood of a sight line intersecting a SNR.
}
\end{enumerate}

It thus appears that X--ray absorption line strengths are primarily
sensitive to galactic and cluster/group properties, and that the systems
that typically host bright AGNs (massive ellipticals in high--density 
environments)
are not conducive to producing absorption like that seen at $z=0$.  In this
case, the Milky Way and Local Group may present a nearly--ideal system
for the production of strong absorption: a
spiral galaxy with sufficient star formation to produce some ionized
gas in the corona, and a low--mass galaxy group that contains a large column
density of intra--group medium at $T\sim 10^6$\,K (but not so massive that this
medium is fully
ionized).  Indeed, even slightly larger galaxy groups may already be
too hot: the poor group of galaxies \citep[containing four \ion{H}{1} 
galaxies;][]{shull98} at $z=0.056$ toward PKS 2155--304
appears to produce only weak \oviii\ absorption \citep{fang02,fang07},
indicating that the oxygen in this system is nearly fully--ionized.

Of course, at this point the above explanation for this discrepancy 
is mostly speculative; quantitative
constraints can only be placed with many more high--S/N X--ray spectra.
The sightlines analyzed herein were chosen solely on the basis of their
being the highest--quality extragalactic \chandra\ grating spectra
available.  Naturally, a search for X--ray absorption in a massive
spiral galaxy hosting a Seyfert--like AGN (or in a quasar sight line
passing close to an intervening spiral) would provide a far more direct
comparison to our local system, but strong AGN outflows 
from the brightest known Seyferts \citep[e.g., NGC 3783;][]{kaspi02}
hinder the detectability of
galactic and intergalactic gas.  A similar
analysis of the far--UV \ovi\ line with \fuse\ is currently in progress,
but this line is expected to trace cooler ($T\sim 10^5$\,K) gas.
To place stronger constraints on the origin of local warm--hot gas,
soft X--ray spectrometers with both large effective areas and high velocity 
resolution ($\la 100$\kms) will be critical
to distinguish between Galactic and extragalactic absorption at $z=0$
and in AGN hosts.  Forthcoming missions such as {\it Constellation--X} and
{\it XEUS} will be able to make great advances in this field. 

\acknowledgments
R.J.W. acknowledges financial support from the Netherlands Foundation for Scientific
Research (NWO).  This work is supported in part by \chandra\ grant AR5--6017X
from the \chandra\ Science Center to S.M.
We are grateful to the \chandra\ team for their efforts on this mission and 
the accompanying library of software and documentation.

\clearpage

\begin{thebibliography}{}
\bibitem[Brighenti \& Mathews(1997)]{brighenti97} Brighenti, F. \& Mathews,
         W.~G.~1997, \apj, 486, L83
\bibitem[Cen \& Ostriker(1999)]{cen99} Cen, R., \& Ostriker, J.~P.
         1999, \apj, 514, 1
\bibitem[Collins et al.(2005)]{collins05}  Collins, J.~A., Shull, J.~M.,
         \& Giroux, M.~L.~2005, \apj, 623, 196
\bibitem[Dav\'e et al.(2001)]{dave01} Dav\'e, R., et al.~2001, \apj, 552, 473
\bibitem[Falomo et al.(1991)]{falomo91} Falomo, R., Giraud, E., Melnick, J.,
         Maraschi, L., Tanzi, E.~G., \& Treves, A.~1991, \apj, 380, L67
\bibitem[Falomo et al.(1993)]{falomo93} Falomo, R., Pesce, J.~E., 
         \& Treves, A.~1993, \apj, 411, L63
\bibitem[Fang et al.(2002)]{fang02} Fang, T.-T., Marshall, H.~L., Lee, J.~C.,
         Davis, D.~S., \& Canizares, C.~R.~2002, \apj, 572, L127
\bibitem[Fang et al.(2003)]{fang03} Fang, T.-T., Sembach, K.~R., \&
         Canizares, C.~R.~2003, \apj, 586, L49
\bibitem[Fang et al.(2007)]{fang07} Fang, T.-T., Canizares, C.~R., \&
         Yao, Y.~2007, \apj, in press (arXiv:0708.1800)
\bibitem[Fields et al.(2007)]{fields07} Fields, D.~L., Mathur, S., Krongold, 
         Y., Williams, R., \& Nicastro, F.~2007, \apj, 666, 828
\bibitem[Kaspi et al.(2002)]{kaspi02} Kaspi, S., et al.~2002, \apj, 574, 643
\bibitem[Kikuchi \& Mikami(1985)]{kikuchi85} Kikuchi, S. \& 
         Yoshitaka, M.~1985, \pasj, 39, 237
\bibitem[Kravtsov et al.(2002)]{kravtsov02} Kravtsov, A.~V., Klypin,
         A., \& Hoffman, Y.~2002, \apj, 571, 563
\bibitem[Martel et al.(2003)]{martel03} Martel, A.~R., et al.~2003,
         \aj, 125, 2964
\bibitem[Mathur et al.(2007)]{mathur07} Mathur, S., 
         et al.~2007, ApJL, Submitted (arXiv:0709.2870)
\bibitem[Nicastro et al.(2002)]{nicastro02} Nicastro, F., et al.~2002,
         \apj, 573, 157
\bibitem[Nicastro et al.(2003)]{nicastro03} Nicastro, F., et al.~2003,
         Nature, 421, 719
\bibitem[Nicastro et al.(2005)]{nicastro05} Nicastro, F., et al.~2005, 
         \apj, 629, 700
\bibitem[Rasmussen et al.(2007)]{rasmussen07}	Rasmussen, A.~P., Kahn, 
         S.~M., Paerels, F., den Herder, J.~W., Kaastra, J., \& de Vries, 
         C.~2007, \apj, 656, 129
\bibitem[Savage et al.(1993)]{savage93} Savage, B.~D., Lu, L., Weymann, R.~J.,
         Morris, S.~L., \& Gilliland, R.~L.~1993, \apj, 404, 124
\bibitem[Savage et al.(2003)]{savage03} Savage, B.~D., et al.~2003,
         \apjs, 146, 125
\bibitem[Sembach et al.(2003)]{sembachetal03} Sembach, K.~R., et al.~2003,
         \apjs, 146, 165
\bibitem[Sembach(2003)]{sembach03} Sembach, K.~R.~2003, preprint
         (arXiv:astro-ph/0311089)
\bibitem[Shull et al.(1998)]{shull98} Shull, J.~M., Penton, S.~V., Stocke,
         J.~T., Giroux, M.~L., van Gorkom, J.~H., Lee, Y.~H., \& Carilli, C.~
	 1998, \aj, 116, 2094
\bibitem[Takei et al.(2007)]{takei07} Takei, Y., Henry, J.~P., Finuguenov,
         A., Mitsuda, K., Tamura, T., Fujimoto, R., \& Briel, U.~G.~2007,
         \apj, 655, 831
\bibitem[Ulrich et al.(1975)]{ulrich75} Ulrich, M.-H., Kinman, T.~D.,
         Lynds, C.~R., Rieke, G.~H., \& Ekers, R.~D.~1975, \apj, 198, 261
\bibitem[Ulrich(1978)]{ulrich78} Ulrich, M.-H.~1978, \apj, 222, L3
\bibitem[Wakker et al.(2003)]{wakker03} Wakker, B.~P., et al.~2003,
         \apjs, 146, 1
\bibitem[Wang et al.(2005)]{wang05} Wang, Q.~D., et al.~2005, \apj,
         635, 386
\bibitem[Williams et al.(2005)]{williams05} Williams, R.~J., et al.~2005, 
         \apj, 631, 856
\bibitem[Williams et al.(2006a)]{williams06a} Williams, R.~J., Mathur, S.,
         Nicastro, F., \& Elvis, M.~2006a, \apj, 642, L95
\bibitem[Williams et al.(2006b)]{williams06b} Williams, R.~J., Mathur, S.,
         \& Nicastro, F.~2006b, \apj, 645, 179
\bibitem[Williams et al.(2007)]{williams07} Williams, R.~J., Mathur, S.,
         Nicastro, F., \& Elvis, M.~2007, \apj, 665, 247
\end{thebibliography}
\end{document}